\documentclass[bibyear]{aa} 

\usepackage{natbib}
\usepackage{graphicx}
\usepackage{txfonts}
\begin{document}

   \title{Deep into the Water Fountains:}
   \subtitle{The case of IRAS 18043$-$2116}

   \author{A.~F. P\'erez-S\'anchez\inst{1,2}\and D. Tafoya\inst{3}\and R. Garc\'{\i}a L\'opez\inst{4}\and W. Vlemmings\inst{3}       
          \and L.~F. Rodr\'{\i}guez\inst{2}
          }

   \institute{European Southern Observatory, Alonso de C\'ordova 3107, Vitacura,
     Casilla 19001, Santiago, Chile. \email{aperezsa@eso.org}
     \and
     Instituto de Radioastronom\'{\i}a y Astrof\'{\i }sica, UNAM, Apdo. Postal 3-72 (Xangari) 58089 Morelia, Michoac\'an, M\'exico.
     \and
     Department of Earth and Space Sciences, Chalmers University of Technology, Onsala Space Observatory, 439 92 Onsala,Sweden.
     \and
     Dublin Institute for Advanced Studies, 31 Fitzwilliam Place, Dublin 2, Ireland.\\
   }



  \abstract
   {The formation of large-scale (hundreds to few thousands of AU) bipolar structures in the circumstellar envelopes (CSEs) of
     post-Asymptotic Giant Branch (post-AGB) stars is poorly understood. The shape of these structures, traced by emission from fast molecular outflows,
     suggests that the dynamics at the innermost regions of these CSEs does not depend only on the energy of the radiation field of the central star.}
   {Multi-frequency observations towards a group of post-AGB sources known as Water Fountain nebulae can help to constrain the nature of
     the mechanism responsible for the launching and collimation of the fast molecular outlflows traced by high-velocity features of H$_{2}$O maser
     emission.}
   {Deep into the Water Fountains is an observational project based on the results of programs carried out with three telescope facilities: The Karl G.
     Jansky Very Large Array (JVLA), The Australia Telescope Compact Array (ATCA), and the Very Large Telescope (SINFONI-VLT).}
   {Here we report the results of the observations towards the WF nebula IRAS 18043$-$2116: Detection of radio continuum emission in the frequency
     range 1.5~GHz - 8.0~GHz; H$_{2}$O maser spectral features and radio continuum emission detected at 22~GHz, and H$_{2}$ ro-vibrational emission lines
     detected at the near infrared.
   }
   {The high-velocity H$_{2}$O maser spectral features, and the shock-excited H$_{2}$ emission detected could be produced in molecular
     layers which are swept up as a consequence of the propagation of a jet-driven wind. Using the derived H$_{2}$ column density, we estimated a molecular mass-loss
     rate of the order of $10^{-9}$~M$_{\odot}$yr$^{-1}$. On the other hand, if the radio continuum flux detected is generated as a consequence of the propagation of a
     thermal radio jet, the mass-loss rate associated to the outflowing ionized material is of the order of 10$^{-5}$~M$_{\odot}$yr$^{-1}$. The presence of a rotating
     disk could be a plausible explanation for the mass-loss rates estimated.}

   \keywords{Stars:AGB and post-AGB -- Masers -- circumstellar matter -- Stars:winds, outflows -- Infrared: stars -- Radio continuum: stars}

   \maketitle
%

\section{Introduction}
Water Fountain (WF) nebulae are thought to represent a subclass of axisymmetric post-Asymptotic Giant Branch (post-AGB) stars
whose collimated outflows are traced by high-velocity ($>100$ km s$^{-1}$) 22 GHz H$_{2}$O maser spectral features. The first WF nebulae was reported by
\citet{Likkel88}. Results from interferometric observations towards the circumstellar envelopes (CSE) of WFs indicate that,
in most cases, the spatial distribution of the detected H$_{2}$O maser components (projected on the plane of the sky) can be correlated with
large-scale bipolar structures \citep[e.g.][]{Sahai052, NaturVle, Gomez11, Lagadec11, Yung11}.

The pumping of the H$_{2}$O maser transitions that trace high-velocity molecular outflows is thought to be
dominated by shocks with neutral molecular gas \citep{Hollenbach13, Gray16}. The pumping mechanism of the H$_{2}$O transitions is very sensitive to changes of
the collisional rates, and requires particular conditions of the density and kinetic temperature of the gas \citep[e.g.][and references therein]{Humphreys2007}. The
detection of high-velocity spectral features of H$_{2}$O masers indicates that there are molecular layers in the bipolar structures where this particular set of
physical conditions are met. Given the velocity of the spectral features detected, the dynamics of these layers might be a consequence of the passage of a
high-velocity outflow throughout the slower and relic AGB wind.

Observational tools such as molecular H$_{2}$ emission at near infrared wavelengths, radio continuum emission, and both Hydroxyl (OH) and H$_{2}$O maser
emission have been used in order to trace the actions of shocks, magnetic fields, and ionized shells on large-scale bipolar structures observed
towards post-AGB stars \citep{Sahai05, Cerrigone08, Walsh09, Bains09, Lagadec11, Cerrigone11, Gledhill12, mio131,Tafoya14, Gonidakis14, Gledhill15}. In
the context of the evolution of the CSEs of late-type stars, multi-frequency observational studies of Water Fountain nebulae provide important information
regarding the physics behind the shapes of the bipolar structures observed towards more evolved counterparts.
In particular, the momentum measured towards bipolar molecular outflows of a sample of 32 post-AGB (or protoplanetary nebulae) stars by \citet{Buja01}
indicates the necessity of an additional energy source (besides radiation pressure) in order to explain the high-velocity components of the CO emission
detected towards these sources. However, the nature of such energy source(s), as well as the parameters related to the high-velocity outflows
(wind velocity, mass in the outflow, magnetic field strength, etc) are yet to be determined.

\section{Framework}

\subsection{H$_{2}$ ro-vibrational lines}

Integral field spectroscopy at near infrared (near-IR) wavelength bands is useful to trace the activity at the innermost region of bipolar post-AGB
sources, for instance, by the detection of molecular Hydrogen (H$_{2}$) emission \citep[e.g.][]{Forde12} . Moreover, the main excitation mechanism of the H$_{2}$ ro-vibrational
lines detected, either UV-pumped or shock-excited, can be constrained using the flux ratios of the detected H$_{2}$ spectral features. In particular, K-band
integral field spectroscopy has yielded important results towards late-type sources at different evolutionary stages, from those young
post-AGB sources ($T_{eff}<1.5\times 10^{4}$~K) which are not hot enough to photoionize the innermost layers of their CSEs \citep{Sanchez08, Gledhill12}, to
the so called ``hot post-AGB'' sources, detected in both near infrared \citep{Gledhill15} and radio wavelengths \citep{Cerrigone08,Cerrigone11}.

In the shock-excited scenario, the populations of the H$_{2}$ ro-vibrational levels depends primarily on both the strength of a passing shock-front, and the density
of the pre-shock region. In this case, the population of the low-vibrational H$_{2}$ levels is favored over that of high-vibrational states. Indeed,
the flux density ratio of the H$_{2}$ ro-vibrational lines 1-0~S$(1)/$2-1~S$(1)\geq 10$ \citep[although values $> 4$ are also considered,][]{Shull78}, is
commonly used as a reliable diagnostic of shock-excited H$_{2}$ emission.

In the radiative scenario, the absorption of Lyman or Werner photons by the
H$_{2}$ molecule leads to a cascade from high- to low-vibrational states via H$_{2}$ ro-vibrational transitions (fluorescence). Hence, UV-pumped emission
eventually lead to the population of both low- and high-vibrational H$_{2}$ states.
In the UV-pumped scenario, the near infrared spectra display both low- and high-vibrational H$_{2}$ ro-vibrational lines,
as well as the Br$\gamma$ and the He I recombination lines, with most of these high-energy lines displaying spatially extended emission. These
lines in particular trace regions with a high degree of ionization which in turn are associated to the strong radiation field of an early-B central star.
In the case of WF nebulae, results of near infrared integral field spectroscopy towards the WF nebulae IRAS 16342$-$3814 were reported by \citet{Gledhill12}. The authors
showed that the spatially resolved H$_{2}$ emission is tracing high-velocity molecular outflows instead of UV-pumped shells. The
non-detection of the Br$\gamma$ emission line, nor He I recombination lines in this spectral window, together with a measured integrated flux ratio
H$_{2}$~1-0~S$(1)/$2-1~S$(1) > 10$, supported a shock-excited scenario for the emission detected.

\subsection{Radio continuum and the Spectral Energy Distribution}

At centimeter radio wavelenghts, the contribution from cold CSE dust and molecular gas is negligible. The flux detected at radio frequencies is
usually associated to free-free interactions in energetic winds \citep{Rey86}. The study of the spectral energy distribution (SED) at radio wavelenghts can
help to constrain the nature of the continuum emission at frequencies where it is expected to be dominated by free electrons (either thermal or non-thermal emission).
The spectral index of the SED and the frequency where the emission becomes optically thin provides information about the density and the temperature distribution
of the emitting particles. For instance, \citet{Bains09} reported the detection of radio continuum emission towards three post-AGB sources. Among them, the SED measured
towards WF IRAS 15445$-$5449 was found to display a steep negative spectral index between $\approx 5$~GHz and $\approx 8.7$~GHz. \citet{mio131} confirmed that
the non-thermal emission is arising from a synchrotron jet, spatially resolved at 22~GHz.

\subsection{IRAS 18043-2116}
The WF IRAS 18043$-$2116 is one of the three hot WF nebulae detected displaying 321~GHz H$_{2}$O maser emission \citep{Tafoya14}.
\citet{Sevenster01} reported on the detection of only one of the two OH main-line transitions (1665 MHz), and both OH satellite lines
(1612 and 1720 MHz) towards this WF source. Indeed, the detection of the 1720 MHz OH maser line is usually associated with shocks and
interacting winds \citep[][and references therein]{Deacon04}.

Based on its position on the MSX and IRAS two-color diagrams, as well as on the line profile of the OH maser lines,  \citet{Sevenster01} associated the
emission to an evolved star, most likely a post-AGB source. \citet{Deacon04} reported on the re-detection of
the OH maser lines reported by \citet{Sevenster01}, and confirmed the non-detection of the 1667 MHz OH maser line. The blue- and red-shifted spectral features of
the double-peaked 1665~MHz OH maser spectrum were detected at 69.7~km~s$^{-1}$ and 104~km~s$^{-1}$, respectively. Therefore, from the velocity of the 1665~MHz OH
maser spectral features, the authors suggested a radial velocity for the stellar source of $87.5$~km~s$^{-1}$.

\citet{Deacon07} reported the detection of high-velocity 22~GHz H$_{2}$O maser emission, with spectral features having velocities outside the velocity
range defined by the double-peaked 1665~MHz OH maser spectrum \citep[$\Delta$v= v$_{red}-$v$_{blue} \approx 33$~km~s$^{-1}$,][]{Deacon04}. This led the authors
to classify the source as a Water Fountain nebulae. \citet{Gonidakis14} reported the detection of both linear and circular polarization of the
OH 1665~MHz maser spectral features. The percentage of circular polarization detected in the strongest spectral feature was $52\%$, which the authors
associated to the $\sigma$-components of Zeeman splitting of this maser spectral line, and consequently, to the action of a magnetic field.\\
Besides the detection of 321 GHz H$_{2}$O maser emission \citep{Tafoya14},  \citet{mio11}, \citet[][2014]{Walsh09} also reported the detection
of the 22~GHz H$_{2}$O maser transition. In most cases the bandwidth used for the observations led to the detection of maser spectral features over
velocity ranges $>~300$~km~s$^{-1}$, a fact that confirmed the results reported by \citet{Deacon07}.

Here we report the first detection of radio continuum emission in the frequency range between 1.0~GHz and 8.0~GHz using the Karl G. Jansky Very Large Array (JVLA) towards
the WF IRAS 18043$-$2116. We also report the detection of the S-and Q-branch H$_{2}$ ro-vibrational lines using SINFONI/VLT for the first time towards
this source. In addition, the result from 2013 observations done using the Australia Telescope Compact Array (ATCA) of both maser and radio continuum
emission at 22~GHz are also presented.

\section{Observations}
\label{sec:Obs}

\subsection{JVLA data}
The observing program (Project code: 15A-301) was executed between June and August 2015, using the most extended (A) configuration of the JVLA, 
in frequency bands L, S, and C. The scheduling blocks were arranged by frequency band. Each one included 11 science sources and their respective flux and phase
calibrators. In this paper we will concentrate on the emission detected towards the WF IRAS 18043$-$2116. The report on the results for the entire sample of targets
will be published in a separate paper.

Each frequency band was observed once during two different days, producing two different data sets of visibilities for each band.
Each data set was calibrated independently. However, both data sets which belong to the same frequency band were eventually combined during the deconvolution
process. An initial calibration of the data was performed using the VLA CASA calibration pipeline. Then, the visibilities were split by source
(science and calibrators) in order to continue the calibration process (when necessary), or to start the image deconvolution process.\\
In the case of WF IRAS 18043$-$2116, the good signal-to-noise ratio (SNR $\approx 60$) within the frequency bands S and C allowed us to calculate the intensity maps
using the multifrequency synthesis algorithm (mfs) mode for wide frequency bands. In our case, the continuum images were calculated over effective bandwidths
of $\approx 1.73$ GHz (S band) and $\approx 1.75$ GHz (C band). In addition, an image of each spectral window was deconvolved in order to measure the flux and
retrieve the spectral energy distribution within both frequency bands (see Table \ref{tbl-1}). On the other hand, the resultant bandwidth at frequency band L
is $\approx$800~MHz. The deconvolution process was performed using the mfs mode, but because the low SNR, it was not possible to deconvolve reliable
maps of individual spectral windows within this frequency band.

\subsection{ATCA Data}
Previous observations towards IRAS 18043$-$2116 were carried out using ATCA in 2013 at 22.2~GHz
(Project code: CX267, Director's Discretionary Time). A bandwidth of 2~GHz centered at rest frequency 22.235~GHz was used for the Compact Array Broadband
Backend (CABB), this using the $32~\times~64$~MHz mode. This setup allowed the implementation of 3 overlaped zoom bands (three 64~MHz channels with 2048 channels
each) centered at rest frequency 22.235~GHz, covering 128~MHz for the maser emission observations (a velocity range of 1725.8~km~s$^{-1}$). The final
spectral resolution for the line observations was 0.42~km~s$^{-1}$. The radio continuum was simultaneously observed using the remaining
bandwidth of 1.8~GHz.

The observing program was executed using the most extended configuration of the ATCA array (6A). The shortest baseline length of this configuration is 337~m
(for comparison, the shortest baseline of the A configuration of the JVLA is 680~m). The calibration of the ATCA continuum data was done using the radio
interferometry data reduction package MIRIAD \citep{Sault95}. It was also used in order to calibrate and restore the H$_{2}$O maser spatial
distribution projected on the plane of the sky. Firstly, the spectral channel with the highest SNR value of the H$_{2}$O maser spectra was
self-calibrated. Then, the solution tables were copied to both set of visibilities, the continuum and the maser visibilities. This allows us to determine the
relative positions between the H$_{2}$O maser spectral features and peak of the radio continuum. In order to determine the projected spatial distribution of
the maser emission, the AIPS task SAD was used on each individual channel of the H$_{2}$O maser data cube. Once the channels with maser emission were
identified, the spatial distribution of the emission on the plane of the sky was determined.

\subsection{SINFONI (VLT) data}

The WF IRAS 18043$-$2116 was observed with the Very Large Telescope at Paranal Observatory, Chile, using the K-band grating of the SINFONI integral field
spectrograph installed at the 8.2-m UT4 \citep{Eisen03, Bonnet04}. The source was observed on 2015 May 10, covering the wavelength range 1.95 - 2.45~$\mu$m with
spectral resolution $\approx 4000$, using the no optics guide star (noAO) mode. The pixel scale was 250~mas which corresponds to a field of view of $8\arcsec \times 8\arcsec$.
The seeing during the observations was $\sim 0.8$~arcsec. The total integration time was 4200~s. Dark, bad-pixel mask, flat-fielding, optical distortion corrections, as well
as wavelength calibration were performed using the SINFONI data-reduction pipeline. In order to correct the spectra for the atmospheric transmission, absorption
features, and flux-calibrate the final datacube, a telluric standard star of spectral type B3V was observed. The telluric standard data were reduced in the same way
as the science source data. The standard star spectrum was extracted from the datacube by averaging the central spaxel spectra of the datacube down to the seeing size. The
standard spectra was then divided by a blackbody at the appropriate temperature. This result was then employed to correct for the telluric
features and atmospheric transmission after the intrinsic features of the standard spectrum were removed. Finally, the continuum substracted images were
constructed by running iteratively the IRAF subroutine CONTINUUM along the dispersion axis at each spatial position across the datacube. The final spaxel resolution
in the spectral direction (channel) obtained is 2.45$\times 10^{-4} \mu$m ($\sim 34.6$~km~s$^{-1}$).

The SINFONI Field-of-View (FoV) was centered at the infrared (IRAS) coordinates reported in the literature for
IRAS 18043$-$2116 ($J2000~18^{\textrm{h}}~7^{\textrm{m}}~21.1^{\textrm{s}}; -21\degr16\arcmin14.2\arcsec$). However, the angular offset of the center of
the SINFONI FoV with respect to the coordinates of the source retrieved from the JVLA and ATCA observations is $\sim 3.95$~arcsec. The absolute uncertainty on the position
of SINFONI-VLT is 1$\arcsec$ - 2$\arcsec$, and the uncertainty on the position of the JVLA observations, as we did not observe using optimized phase referencing calibrators,
should be of the same order. With these uncertanties, the radio source and the source at the center of the SINFONI FoV cannot be associated. The
coordinates of the radio source from our JVLA observations are $(J2000)~18^{\textrm{h}}~7^{\textrm{m}}~20.853^{\textrm{s}}; -21\degr16\arcmin12.271\arcsec$. The closest
source to this position within the SINFONI FoV has an angular offset of 1.82$\arcsec$. Although it is detected towards the edge of the SINFONI FoV, we assume that the emission
detected towards this position is associated to our science target. A spectrum of this feature was extracted by averaging 3$\times$3 pixels at the detected position
along the dispersed axis.

\section{Results}

\subsection{Radio continuum emission}   

   \begin{figure*}
   \centering
   \includegraphics[width=18.0cm]{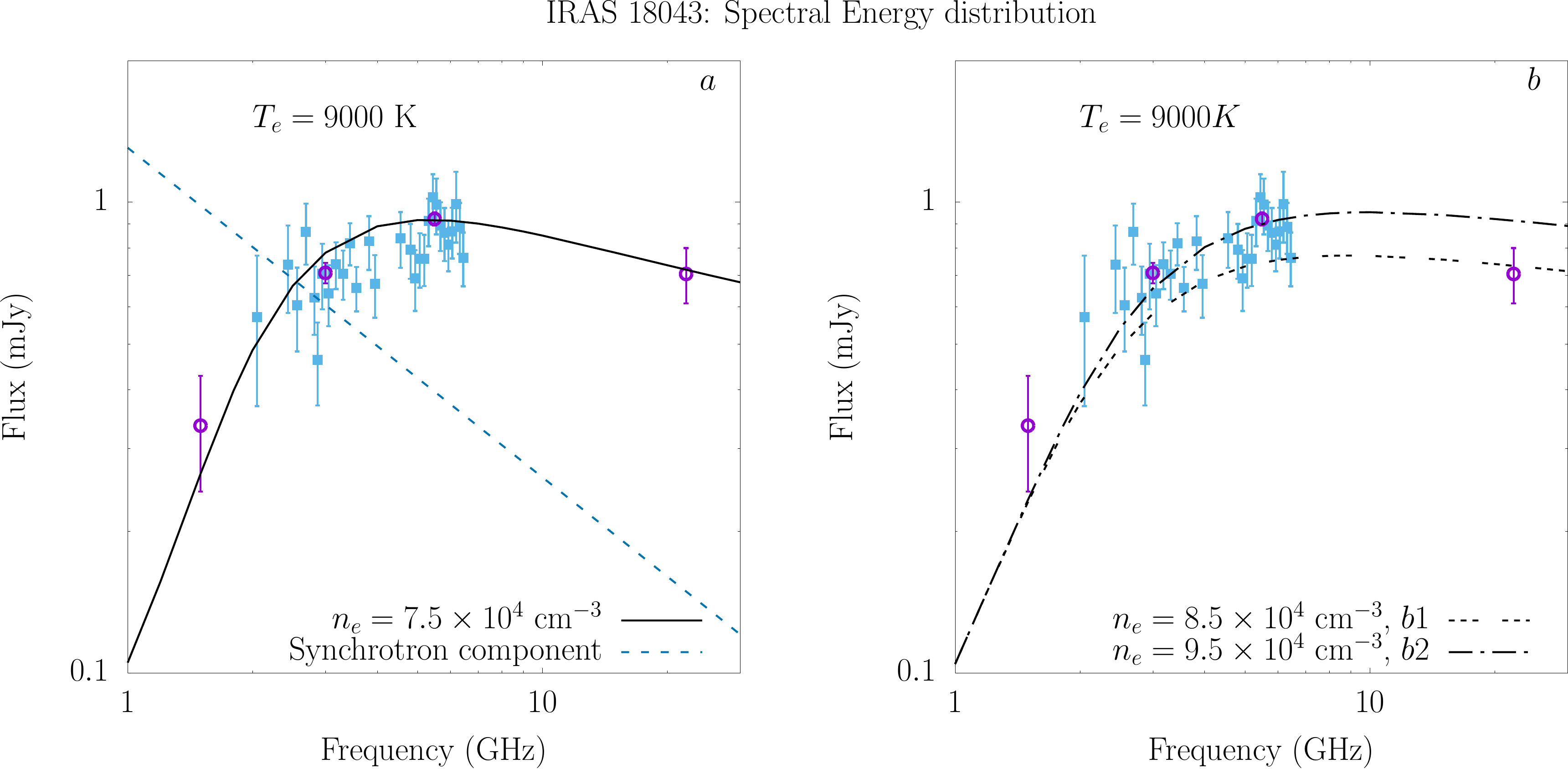}
   \caption{Spectral energy distribution measured towards the WF IRAS 18043$-$2116. The small filled squares are the measurements
  at individual spectral windows within JVLA frequency bands C and S. The open circles represent the fluxes obtained from mfs evaluated
  at the central frequency of JVLA bands L, S, and C. The flux at 22~GHz was measured in 2013 with ATCA. The different lines in the panels represent the fit
  to the data points obtained from radiative transfer of radio continuum emission (see text). In panel {\it a} the model includes both thermal and non-thermal components,
  assuming the latter is synchrotron radiation with flux at 22~GHz of 0.15 mJy and spectral index $=-0.7$ (dashed line). At low frequencies the synchrotron component is
  mostly absorbed by optically thick free-free and does not contribute to the total emission. In panel {\it b}, the
  models consider thermal emission only. Model $b1$ fits the flux at 22~GHz measured with ATCA in 2013, whereas Model $b2$ fits the
  fluxes measured with JVLA in 2015. The difference between both thermal models suggests an increase of the 22~GHz flux density of $\approx~$20\%
  between 2013 and 2015.\label{fig:sed}}%
    \end{figure*}

%
\begin{table}[!h]
\caption{Radio continuum fluxes towards the WF IRAS 18043-2116.} 
\label{tbl-1}      
\centering                          
\begin{tabular}{c | c c | c c | c}        
\hline\hline                 
Frequency& & \multicolumn{2}{c}{Intensity (mJy)} & & Beam \\    
(MHz) & \multicolumn{2}{c|}{2.0~GHz band} & \multicolumn{2}{c|}{0.1~GHz spw} & (arcsec$^{2}$)\\ \cline{2-5}
& S$_{\nu}$ & rms & S$_{\nu}$ & rms &\\
\hline                        
   1498.994  & 0.335 & 0.093 &&  & 3.60$\times$0.96\\
   2051.006  & & & 0.570 & 0.201 & 1.54$\times$0.82\\
   2435.007  & & & 0.737 & 0.155 & 1.34$\times$0.70\\
   2563.008  & & & 0.604 & 0.122 & 1.26$\times$0.66\\
   2691.008  & & & 0.865 & 0.128 & 1.21$\times$0.62\\
   2819.008  & & & 0.627 & 0.104 & 1.17$\times$0.60\\
   2947.009  & & & 0.705 & 0.112 & 1.11$\times$0.58\\
   2999.012  & 0.708 & 0.036 & & & 0.93$\times$0.48\\
   3051.009  & & & 0.640 & 0.095 & 1.11$\times$0.56 \\
   3179.009  & & & 0.738 & 0.084 & 1.07$\times$0.55\\
   3307.010  & & & 0.705 & 0.084 & 1.02$\times$0.53\\
   3435.010  & & & 0.818 & 0.084 & 0.98$\times$0.51\\
   3563.011  & & & 0.658 & 0.071 & 0.96$\times$0.49\\
   3819.011  & & & 0.826 & 0.108 & 0.88$\times$0.45\\
   3947.012  & & & 0.671 & 0.102 & 0.88$\times$0.45\\
   4551.006  & & & 0.839 & 0.114 & 0.71$\times$0.39\\
   4807.006  & & & 0.793 & 0.106 & 0.69$\times$0.38\\
   4935.006  & & & 0.690 & 0.102 & 0.66$\times$0.36\\
   5063.007  & & & 0.759 & 0.100 & 0.64$\times$0.35\\
   5191.007  & & & 0.758 & 0.095 & 0.63$\times$0.34\\
   5319.007  & & & 0.912 & 0.106 & 0.61$\times$0.33\\
   5447.007  & & & 1.026 & 0.121 & 0.60$\times$0.33\\
   5499.008  & 0.922 & 0.030 & & & 0.54$\times$0.30\\
   5551.007  & & &  0.989 & 0.134 & 0.58$\times$0.32\\
   5679.007  & & &  0.893 & 0.106 & 0.58$\times$0.31\\
   5807.008  & & &  0.862 & 0.111 & 0.57$\times$0.31\\
   5935.008  & & &  0.813 & 0.100 & 0.54$\times$0.30\\
   6063.008  & & &  0.866 & 0.107 & 0.52$\times$0.30\\
   6191.008  & & &  0.991 & 0.170 & 0.47$\times$0.28\\
   6319.008  & & &  0.887 & 0.110 & 0.50$\times$0.28\\
   6447.008  & & &  0.763 & 0.100 & 0.49$\times$0.29\\
   22227.259 & 0.705 & 0.095 & &  & 3.0$\times$0.53\\
\hline                                   
\end{tabular}
\end{table}

The fluxes measured at the different spectral windows of frequency bands L, S and C of the JVLA, as well as the flux measured with ATCA at 22~GHz, are listed in
Table \ref{tbl-1}. All the fluxes were obtained from fitting a 2-dimensional Gaussian model to the (spatially unresolved) brightness distribution in the restored
images using the AIPS task {\it JMFIT}. The uncertainty of the measurements correspond to the 1-$\sigma$ error of the Gaussian model. The spectral energy distribution (SED) is
presented in Fig. \ref{fig:sed}. The flux measured within band C is higher than the fluxes measured in bands S and L, where the values of the flux density
are $\approx$23$\%$ and $\approx$64$\%$ lower, respectively. Moreover, the flux measured with ATCA at 22~GHz in 2013 is $\approx$24$\%$ lower than the flux measured
in band C. Although the source is not spatially resolved at centimeter wavelengths, it is worth noting that the extent of the shortest baseline of the
ATCA 6A array configuration is about half the extent of the shortest baseline of the JVLA A configuration. This allow us to rule out the possibility of flux losses
due to resolving out extended structures of the source at 22~GHz.

Overall, the SED of IRAS 18043$-$2116 resembles the radio continuum spectrum expected from stellar winds \citep{Rey86}. The shape of the SED indicates that the
radio continuum emission is optically thick below $\approx$5.5~GHz, with slope $\alpha <+2$. Above 5.5~GHz the slope turns over into a flat, optically thin spectrum.
Using the value of the flux measured with ATCA at 22~GHz to calculate the spectral index above the turnover frequency ($\nu_{c}=5.5$~GHz) we obtained
$\alpha= -0.2 \pm 0.2$. This could suggest the presence of a non-thermal component which significantly contributes to the radiation
field detected. However, this value must be carefully considered as there is a gap of two years between the JVLA and the ATCA observations.
Indeed, radio continuum detected towards post-AGB stars have shown variability on time-scales from a few to tens of years \citep{Cerrigone11}. Nonetheless, the
optically thin nature of the SED above 5.5~GHz can still be inferred from the fluxes measured at the high frequency spectral windows within band C
(see Fig. \ref{fig:sed}).

The main constraint on the interpretation of the data is the flux density measured at 22 GHz. As metioned above, the measurement of the flux at 22~GHz was
done two years before the JVLA observations, therefore it might have changed. In Fig. \ref{fig:sed} we present two plausible scenarios
based on the best fits of our data sample. In the first scenario (panel {\it a}) we assumed that the flux at 22~GHz did not change since the measurements of 2013. In this
scenario we can fit the data with a model that considers both, thermal and non-thermal components. For such a model we use a synchrotron (non-thermal component) flux
at 22~GHz of 0.15 mJy and spectral index $\alpha_{s}=-0.7$ (Dashed line in panel {\it a}). Moreover, in the model the synchrotron source is surrounded
by a thermal electron sheath with density $n_{e}=$7.5$\times$10$^{4}$~cm$^{-3}$ and temperature $T_{e}=9000$~K, which is responsible for the thermal component
of the emission detected.\\
The second scenario is based on the variability of the flux density at 22~GHz. In this case we assume that the flux might have increased
rather than decreased between 2013 and 2015. Although the latter cannot be ruled out, a lower flux density at 22~GHz in
2015 might imply the presence of a stronger synchrotron (non-thermal) component.
In Fig \ref{fig:sed}, panel {\it b} shows the results of the radiative transfer  models including only thermal emission, assuming the source is an
ionized shell. Model $b1$ (dotted line) is based on the flux density measured with ATCA at 22~GHz, and therefore underestimates the fluxes measured in 2015 within the JVLA
frequency bands. The electron density and temperature values obtained from the model are $n_{e}=8.5\times10^{4}$~cm$^{-3}$ and $T_{e}=9000$~K.\\
Model $b2$ (dash-dot line) is based on the fluxes measured with JVLA in 2015. The best fit indicates that the flux density at 22~GHz may have increased about
20\% between 2013 and 2015. In this case, the electron density and temperature retrieved from the model are $n_{e}=9.5\times10^{4}$~cm$^{-3}$ and T$_{e}=9000$~K.
Therefore, the thermal emission models suggest that the electron density in the emitting region should have increased by $\approx$10\% between 2013 and 2015.
Nontheless, both scenarios could be related to the propagation of a shock-front throughout a denser circumstellar envelope \citep[][and references
therein]{Cerrigone11}.

\subsection{High-velocity H$_{2}$O maser emission}   
   
\begin{figure}
   \centering
   \includegraphics[width=9.5cm]{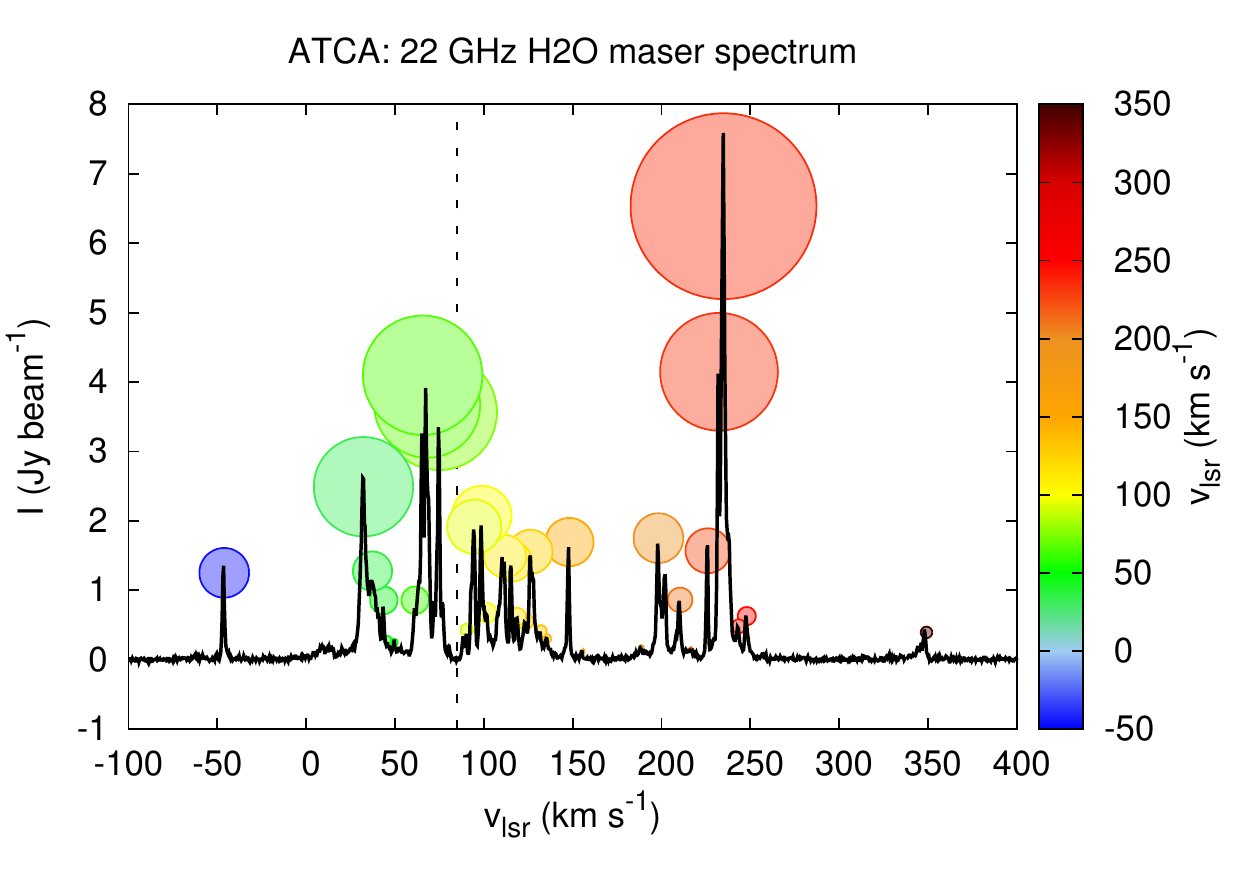}
   \caption{The 22~GHz H$_{2}$O maser spectrum detected towards the WF IRAS 18043$-$2116. The systemic velocity derived for the stellar source from the OH maser spectrum
  is v$_{lsr,\star}=87.5$~km~s$^{-1}$ \citep[vertical dashed line,][]{Deacon04}. The final spectral resolution is $\Delta v=0.42$ km~s$^{-1}$, and the
  total velocity range covered by the maser spectral features is 400 km~s$^{-1}$.\label{fig:spectrum}}
   \end{figure}

Based on the spectral features of the 1665~MHz OH maser line, \citet{Deacon04} suggested a systemic velocity for the stellar source of
v$_{\star,lsr}= 87.5$~km~s$^{-1}$ (dashed line in Fig. \ref{fig:spectrum}). The H$_{2}$O maser spectrum obtained with ATCA is presented
in Fig. \ref{fig:spectrum}. The spectral features of the H$_{2}$O maser emission were detected spread over a velocity range of 400~km~s$^{-1}$. We identified
33 H$_{2}$O maser spectral features that were color-coded as blue- and red-shifted according to their velocity with respect to v$_{\star,lsr}$. 
The extreme spectral features appear at LSR velocities of 349~km~s$^{-1}$ (red-shifted) and $-$46~km~s$^{-1}$ (blue-shifted). In total, the blue-shifted emission covers a velocity
range of 132~km~s$^{-1}$, with the (blue-shifted) brightest features moving at 20~km~s$^{-1}$ with respect to v$_{\star,lsr}$. On the red-shifted side, the spectral
features spread over a wider velocity range (264~km~s$^{-1}$). The brightest maser spectral feature in the spectrum arises from the red-shifted side, moving with
velocity along the line-of-sight of 150~km~s$^{-1}$ with respect to v$_{\star,lsr}$.

\begin{figure}
   \centering
   \includegraphics[width=9.5cm]{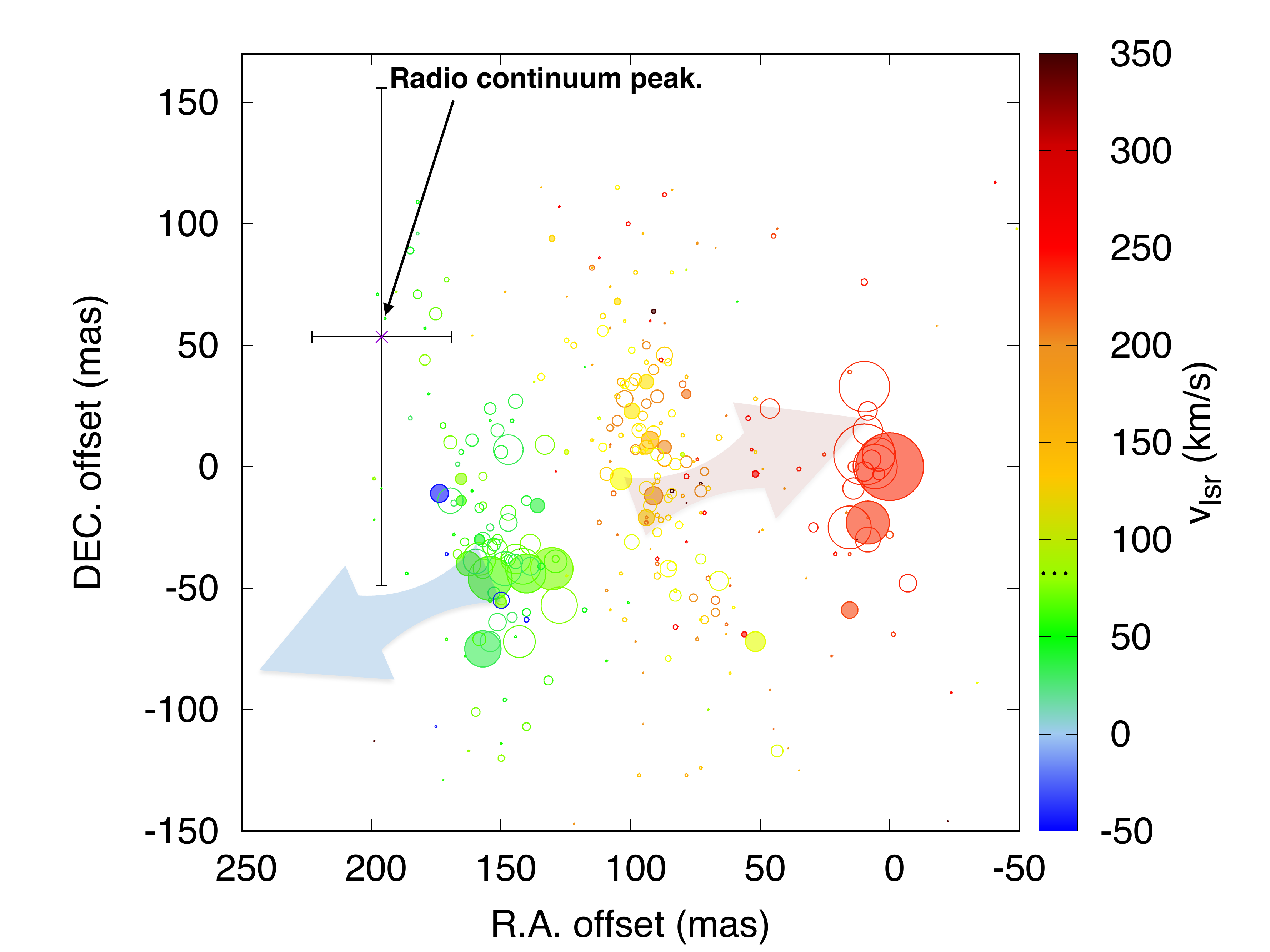}
    \caption{Spatial distribution of the radio continuum and the spectral features of H$_{2}$O maser emission detected at 22~GHz with ATCA. Both the radio
    continuum and the H$_{2}$O maser observations were carried out simultaneously in 2013. The star with error bars shows the relative
    position of the peak of radio continuum detected at 22~GHz with respect to the brightest H$_{2}$O maser feature. The size of
    each circle scales linearly with the intensity of emission measured in each channel related to H$_{2}$O spectral features. Each filled circle represents the
    brightest channel of each spectral figure in Fig. \ref{fig:spectrum}. The dots in the color bar indicate the velocity of the stellar source
    v$_{\star,lsr}$. The blue and red arrows indicate a possible orientation (projected on the plane of the sky) of the axis defined by the high-velocity
    outflow traced by the H$_{2}$O maser emission.\label{fig:distribu}}
   \end{figure}

The projected spatial distribution of the H$_{2}$O maser emission is presented in Fig. \ref{fig:distribu}. The size of the circles scales as the peak intensity value measured
at each individual channel. The offsets in this figure are with respect to the position of the brightest spectral feature. The relative position of the radio continuum
emission detected at 22~GHz is also shown in Fig. \ref{fig:distribu}. The position uncertainty of the maser features is a function of both the synthesized beam $\theta_{22~GHz}$
and of the signal-to-noise ratio (SNR) measured at each channel of the spectrum; this is
$\Delta\theta=0.45 \frac{\theta_{22GHz}}{SNR}$ \citep{Reid88}. The peak flux of both, the strongest and one of the weakest spectral features in
Fig \ref{fig:spectrum}, as well as the rms value measured at the source position in a line-free channel (rms=$1.02\times 10^{-2}$~Jy/beam) were used in order to estimate
the SNR. Thus, the minimum and maximum positional uncertainties along the major and minor axis of the synthesized beam for the maser observations
($2.16\arcsec$, $0.43\arcsec$, $18.37\degr$) are respectively 1.71~mas$\times$0.34~mas and 60~mas$\times$12~mas. Therefore, taking into account the
positional error of the weak spectral features, all the H$_{2}$O maser features arise from within 180~$\times$~200~mas$^{2}$ projected on the plane of
the sky. The blue- and red-shifted emission appear arising from two separated groups. The spatial distribution of the blue-shifted emission suggests that all the
spectral features arise within the same line-of-sight. In turn, the red-shifted emission is generated in two regions with angular separation of 90~mas from each other,
both towards the East from the blue-shifted emission. The projected position of the brightest maser spectral features suggests that it is generated in a
region separated from the bulk of the red-shifted emission.

\subsection{SINFONI K-band Integral field spectroscopy}

\begin{table}
\caption{H$_{2}$ ro-vibrational lines detected towards the WF IRAS 18043-2116.}
\label{tbl-2}
\centering
\begin{tabular}{c|cccc}
  \hline\hline
  Line & $\lambda_{0}$ & $\lambda_{p}$ & F & $\Delta$F\\
       & ($\mu$m) & ($\mu$m) & \multicolumn{2}{c}{$\times$10$^{-16}$ erg~cm$^{-2}$~s$^{-1}$}\\ 
\hline
1 0 S(3) &   1.9575564 & 1.95749 & 5.45 & 0.23\\
1 0 S(1) &   2.1218299 & 2.12199 & 6.22 & 0.14\\
2 1 S(2) &   2.1542257 & 2.15393 & 0.75 & 0.097\\
1 0 S(0) &   2.2233001 & 2.22343 & 2.15 & 0.24\\
2 1 S(1) &   2.2477228 & 2.24783 & 1.06 & 0.23\\
1 0 Q(1) &   2.4065935 & 2.40666 & 5.51 & 0.37\\
1 0 Q(2) &   2.4134354 & 2.41376 & 3.38 & 0.53\\
1 0 Q(3) &   2.4237289 & 2.42398 & 5.30 & 0.43\\
1 0 Q(4) &   2.4374914 & 2.43769 & 4.56 & 1.11\\
\hline
\end{tabular}
\end{table}

   \begin{figure*}
   \centering
   \includegraphics[width=18.0cm]{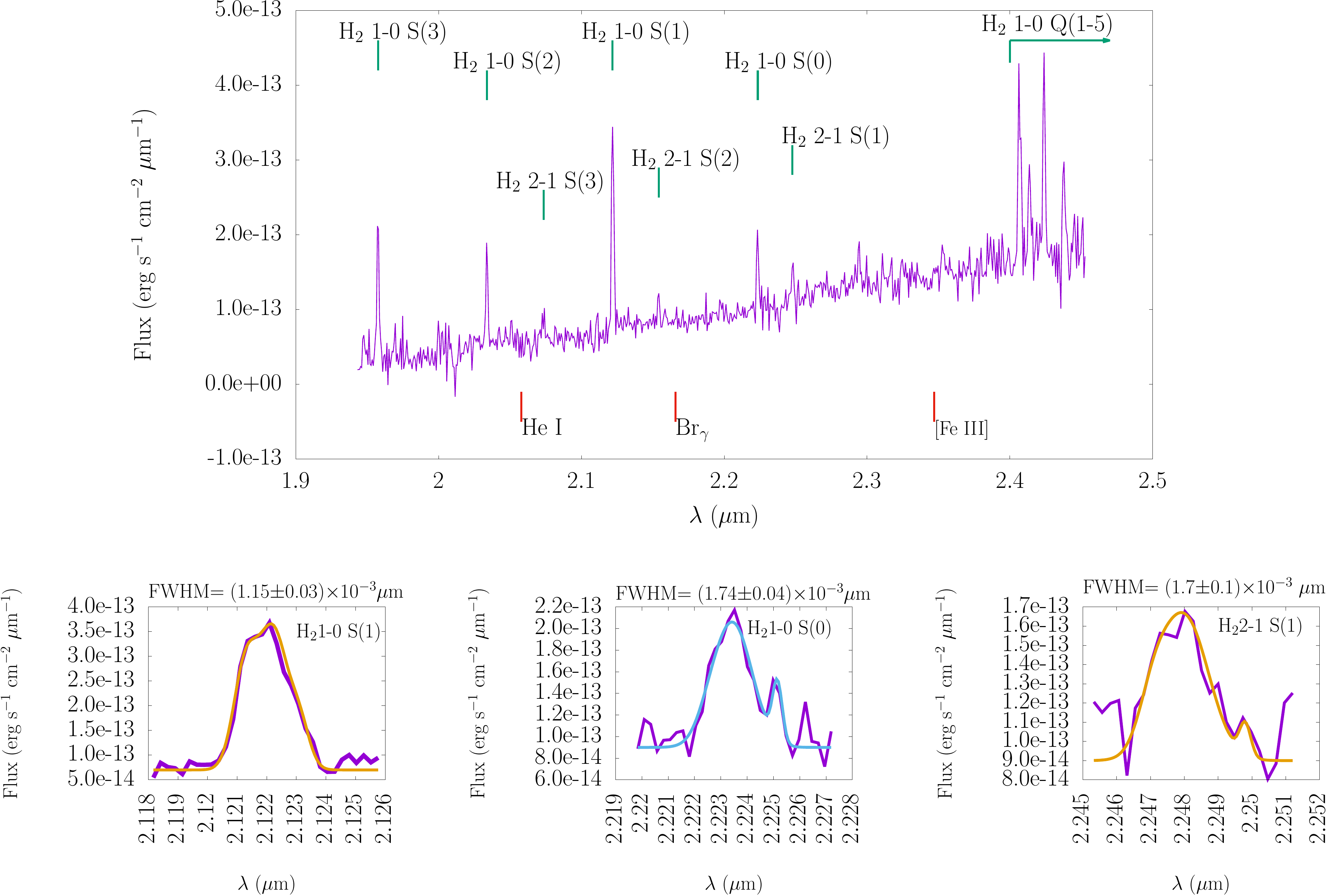}
   \caption{\label{fig:NIRspectrum} SINFONI K-band spectrum detected towards the WF IRAS 18043$-$2116. Both, Q- and low energy S-branch H$_{2}$ ro-vibrational
    transitions dominate the spectra. Non-energetic H$_{2}$ ro-vibrational lines, neither Br$_{\gamma}$ emission lines nor He recombination lines, were detected.
    The bottom panels display the ro-vibrational lines H$_{2}$ 1-0 S(1), H$_{2}$ 1-0 S(0), H$_{2}$ 2-1 S(1). A simple sum of two gaussian functions was
    implemented in order to fit and then estimate the FWHM of each spectral feature.}%
    \end{figure*}

The spectrum obtained with SINFONI (K-band: 1.95$\mu$m - 2.45 $\mu$m) is shown in Fig. \ref{fig:NIRspectrum}. The spectral features detected correspond
to the 1-0 Q(1-4)-branch, as well as to low-vibrational S-branch lines. The fluxes measured in the different H$_{2}$ lines detected are listed in
Table \ref{tbl-2}. None high-excitation lines such as He I recombination lines nor the Br$\gamma$ line were detected. The steep positive spectral
index of the continuum baseline within the SINFONI K-band is indicative of a reddened source, likely due to the presence of a dusty structure where the
stellar source is still embedded.

The extinction can be estimated using the line flux ratio 1-0Q(3)/1-0S(1). Because the strength ratio of these transitions is
intrinsic and fixed to their respective spontaneous decay rate ratio ($R_{0} = 0.70$), any other value would be an indicator of extinction. From the fluxes in
Table \ref{tbl-2} we obtain 1-0Q(3)/1-0S(1)$= 0.85 \pm 0.01$. Using the Rieke \& Lebofsky (1985) extinction law, we estimated
A$_{V}=5$~mag. The ratio 1-0S(1)/2-1S(1)~$\geq 4$ is used as a reliable diagnostic of shock-excited H$_{2}$ emission
\citep{Shull78}. Using the fluxes in Table \ref{tbl-2}, we obtain 1-0(S1)/2-1(S1)$= 5.86 \pm 1.27$. This value
should increase if any line-of-sight extinction correction is applied to the fluxes measured. This fact supports the shock-excited
origin of the ro-vibrational lines detected.

The Full Width at Half Maximum (FWHM) of most of the H$_{2}$ ro-vibrational lines detected are, in average, a factor of 5 - 6 larger than the final spectral resolution unit
obtained in the SINFONI K-band. In Fig. \ref{fig:NIRspectrum} we present the estimated FWHM for three low-vibrational lines of the spectrum. A simple 1D gaussian
model was implemented in order to obtain standard dispersion and then to compute the FWHM$~= 1.15\times 10^{-3}, 1.69\times 10^{-3}, 1.74\times 10^{-3}$~$\mu$m for
the H$_{2}$ lines 1-0S(1), 1-0S(0), 2-1S(1), respectively. The spectral resolution unit of our spectrum is $\Delta\lambda=2.45\times 10^{-4}$~$\mu$m
($\Delta$v$= 34.6$~km~s$^{-1}$). Consequently, the velocity ranges defined by the blue and red wings of the spectral lines at the bottom in Fig \ref{fig:NIRspectrum} are,
respectively, 163~km~s$^{-1}$, 246~km~s$^{-1}$, and 239~km~s$^{-1}$. The linewidth of shock-excited H$_{2}$ ro-vibrational lines is a function of the inclination of the
axis defined by the shock front with respect to the line-of-sight.  The estimated velocity values agree with the velocities traced by the high-velocity spectral features
of H$_{2}$O maser emission. However, high-spatial resolution observations are crucial in order to probe the spatial correlation between the gas traced by the high-velocity
emission of H$_{2}$, and the gas traced by the high-velocity spectral features of H$_{2}$O maser emission.

\begin{figure}
   \centering
   \includegraphics[width=9.3cm]{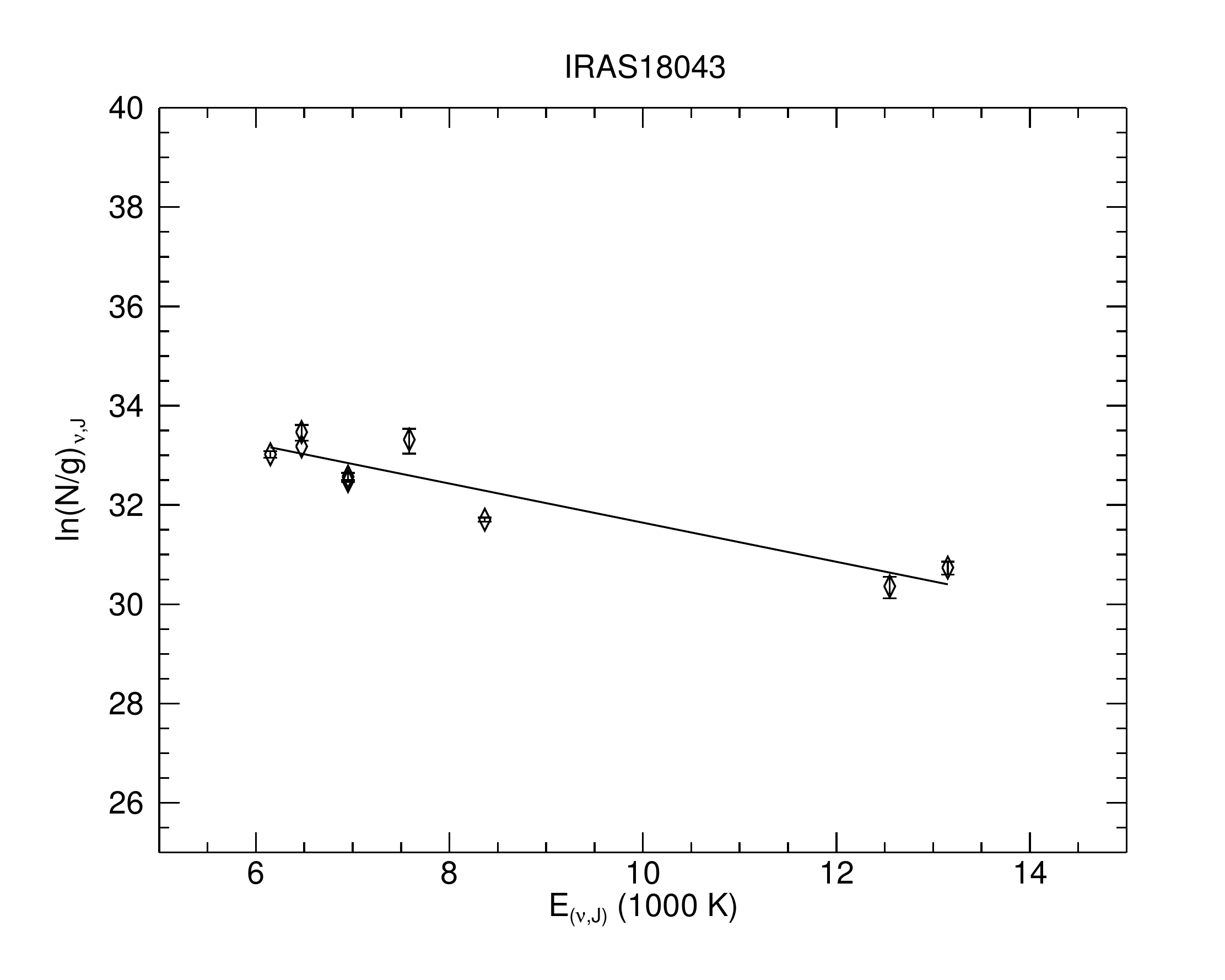}
      \caption{H$_{2}$ rotational diagram relating the column density (N$_{H_{2}}$) of the upper level of the detected transitions with their
  respective excitation temperature in thermal equilibrium. The gas temperature estimated from the best fit to the data is T$_{g} = 2538 \pm 393$~K,
  and the estimated column density is N$_{H_{2}} = 2.348\times 10^{-17}$cm$^{-2}$. \label{fig:excitadiagram}}
   \end{figure}

The H$_{2}$ rotational diagram in Fig. \ref{fig:excitadiagram} shows the logarithm of the column density (N$_{\nu,J}$) of the upper
ro-vibrational state (divided by its statistical weight g$_{\nu,J}$) of the different H$_{2}$ transitions as a function of their excitation energy.
Assuming thermal equilibrium, the slope of the best model fitting the data points is inversely proportional to the temperature of the molecular (H$_{2}$) gas.
Using the H$_{2}$ line fluxes measured, the column density of the emitting region is N$_{H_{2}} = 2.35\times 10^{17}$cm$^{-2}$ and the
estimated gas temperature is T$_{g}=2538 \pm 393$~K. In particular, the temperature estimated from the H$_{2}$ rotational diagram agrees with the
kinetic temperature of the gas required in order to generate not only the 22~GHz line, but also higher H$_{2}$O maser transitions such as the 321~GHz H$_{2}$O
maser emission detected by \citet{Tafoya14} towards this source.

\section{Discussion}
\label{sec:discusions}

\subsection{Spatial scale traced by the H$_{2}$O maser emission}

The spatial distribution of the maser emission presented in Fig. \ref{fig:distribu} agrees with the results presented by Walsh et al. (2014). The overlap between the
projected positions of the extreme velocity spectral features and the spectral features with LSR velocities close to v$_{\star,lsr}$, suggest that
the gas traced by the H$_{2}$O maser emission propagates nearly along the line-of-sight (Walsh et al. 2009). Furthermore, the high-velocity range covered by both,
the H$_{2}$ ro-vibrational lines and the H$_{2}$O maser spectral features indicates that the emission originates from molecular gas moving with velocities of
hundreds of km~s$^{-1}$ with respect to v$_{\star,lsr}$.

The shock-excited ro-vibrational H$_{2}$ lines detected with SINFONI-VLT are most likely tracing the activity close to the region where the high-velocity winds
are launched. Given that the main pumping mechanism of the H$_{2}$O maser transition is thought to be collisional, via neutral
molecules such as H$_{2}$, a direct link could be drawn between the regions traced by the detected H$_{2}$ and H$_{2}$O spectral features. However, the
H$_{2}$O maser emission might trace outer molecular layers along the high-velocity outflows where the inversion of the level population is not quenched by
high collisional rates.
Assuming that the molecular gas pumping the H$_{2}$O maser transition is driven by the high-velocity outflows, and taking into account the error in the position of
the weak H$_{2}$O maser spectral features, the spatial distribution of the maser spectral features suggests that
the mechanism driving the high-velocity outflow is one of the main (if not the most important) inputs of energy for the dynamics
in regions with areas of a few thousand AU$^{2}$. New high angular resolution observations at near-IR and radio frequencies, improving the positioning of the source on
the field of view at near-IR, could in principle probe the spatial distribution of both, the shock-excited ro-vibrational lines and the maser spectral features; and therefore
confirm our interpretation.

\subsection{Radio continuum flux: Radio jet or HII region?}

The relative position of the radio continuum peak detected with ATCA at 22~GHz is shown with respect to the brightest H$_{2}$O spectral feature in Fig. \ref{fig:distribu}, and
the nature of the source of the radio continuum emission can be inferred from the SED in Fig. \ref{fig:sed}. The spectral energy distribution indicates
that the flux is dominated by a thermal component, with turnover frequency at 5.5~GHz. For reference, the SED from an
idealized HII region (a homegeneous, uniform, isothermal, ionized wind) results in $\alpha = +2$ in the optically thick regime, whereas above the
turnover frequency, the emission becomes optically thin and $S_{\nu}\propto \nu^{-0.1}$ \citep[e.g.][and references therein]{Kurtz05}. In these cases the radio
continuum is associated to Bremsstrahlung emission from the interaction of free electrons and ions in photoionized regions around hot (T$_{eff}> 2.0 \times 10^{4}$~K) central
radiation sources. The interpretation of the results of radiative transfer models presented in Fig. \ref{fig:sed} depends on whether the flux density measured at 22~GHz
remained stable during the period between 2013 and 2015. If the flux density did not change, our best model indicates that the radio continuum emission, although dominated
by a thermal component, has a small but significant synchrotron (non-thermal) component with flux density of 0.15~mJy at 22~GHz. This scenario implies the presence of a strong
magnetic field which could be related to the collimation of the high-velocity outflow traced by the H$_{2}$O maser emission \citep[e.g.][]{mio131}. Nonetheless, new
observations of radio continuum emission above 10 GHz are needed in order to probe the source of the non-thermal component.

Assuming that the flux density detected at 22~GHz has increased between 2013 and 2015, our best fits indicates that the source of the radio continuum
might be Bremsstrahlung emission from thermal electrons. Our model suggests that an increase of the electron density of $\approx$~10\% between 2013 and 2015
is needed in order to fit the SED measured from the JVLA data taken in 2015. Such increase of the electron density in the emitting region can be associated to
the propagation of an ionizing shock front from the innermost region of the relic AGB circumstellar envelope. Note that for both scenarios, the models
underestimate the flux density at lower frequencies, which suggest that the spectral index in the optically thick regime of the SED is $\alpha < +2$. Indeed,
spectral index values $\alpha \leq +1.5$ in the optically thick regime, and the gradual change of the slope near the turnover frequency, are usually related to
inhomogeneous (non-spherical), jet-like stellar winds \citep{Rey86}.

Using the velocity v$_{\star,lsr}$ derived from the double-peak 1665~MHz OH maser spectra for the WF IRAS 18043$-$2116 \citep{Deacon04} and the revised kinematic distance
calculator \citep{Reid09}, we estimated both near and far kinematic distances: $6.2\pm 0.3$~kpc and $10.4\pm 0.3$~kpc, respectively. Hence, from the flux measured at the
optically thin regime of the SED, the number of ionizing photons per second ($\dot{N}_{i}$) needed in order to keep ionized the region detected in radio continuum can
be estimated using the following relation \citep{Schraml}:
\begin{equation}
  \dot{N}_{i} = 8.9\times 10^{43} \Big(\frac{S_{\nu}}{\textrm{mJy}}\Big)\Big(\frac{\nu}{4.9~\textrm{GHz}}\Big)^{0.1}\Big(\frac{D}{\textrm{kpc}}\Big)^{2}\Big(\frac{T_{e^{-}}}{10^{4}\textrm{K}}\Big)^{-0.45}~s^{-1}.
  \label{eq:numberphot2}
\end{equation}

Then, with S$_{5.935{\textrm{\tiny{GHz}}}}= 0.81$~mJy (see Table \ref{tbl-1}) in Eq. \ref{eq:numberphot2} and T$_{e}=9000$~K, we obtain 
$\dot{N}_{i} = 8.37\times 10^{45}$ s$^{-1} (D/10.4$ kpc$)^{2}$. Despite the given dependence on the distance, the number of ionizing photons
per second calculated is of the same order of magnitude as the flux of Lyman continuum photons that stars with spectral type B1 or B2 (III) can
generate \citep{Panagia73}. This, in principle, would be compatible with assuming that the radio continuum emission arises from an ionized region surrounding a hot stellar
source associated to the WF IRAS 18043$-$2116, whose temperature must be T$_{eff} > 17000$~K, as is the case of hot post-AGB souces \cite[][]{Cerrigone08,Cerrigone11,Gledhill15}.
Nevertheless, the FWHM values estimated for the H$_{2}$ lines detected in our SINFONI K-band spectrum indicate that the
broadening of the line is larger than what could be expected from Doppler shifting in a radiative-excited scenario. In addition, the lack of high-energy
H$_{2}$ lines, as well as the non-detection of Br$\gamma$ and He recombination lines, also support a shock-excited scenario over the UV-pumped scenario
for the H$_{2}$ emission. Hence, although the radiative scenario cannot be ruled out, the radio flux detected is most likely the result of the propagation
of a (partially) ionizing shock front, generated by the interaction between a collimated high-velocity outflow and the steady-expanding and relic CSE formed in the AGB
phase. The molecular layers swept up by the shock front might be related to those traced by the H$_{2}$O maser and H$_{2}$ line emission.
Nonetheless, regardless the source of the emission, a turnover frequency, feature rarely observed towards low- and intermedate-initial mass late-type stars, can still be inferred from the JVLA data.

\subsection{Velocity field and Mass-loss rates}

The FWHM of the H$_{2}$ lines is consistent with the velocity range covered by the H$_{2}$O maser features,
where the most blue- and red-shifted spectral features appear moving respectively at 132~km~s$^{-1}$ and 264~km~s$^{-1}$ with respect to the systemic
velocity of the source (see Fig. \ref{fig:spectrum}). If the direction of propagation of the bulk of the shocked gas is not along the line-of-sight, then the
velocity measured for the H$_{2}$ and H$_{2}$O will be only a fraction of the total velocity of the high-velocity outflow.

Using the column density derived from the H$_{2}$ rotational diagram (Fig. \ref{fig:excitadiagram}), a ``mass-loss rate'' value can be associated to the H$_{2}$
molecular gas component, shock-excited and driven by the passage of a high-velocity outflow ($\dot{M}_{H_{2}}$). This value can be estimated using:

\begin{equation}
\dot{M}_{H_{2}}=2\mu m_{H}N_{H_{2}}A\frac{dv_{T}}{dl_{p}},
\label{eq:otraeq}
\end{equation}
\noindent

\noindent where $\mu$ is the mean atomic weight, m$_{p}$ is the proton mass, and A is the emitting area, $N_{H_{2}}$ is the column density,
$dv_{T}$ is the tangential component of the velocity vector of the high-velocity outflow, and $dl_{p}$ is the projected length of the emitting region. Because the
emitting source is not spatially resolved, the projected length of the emitting region as well as its area are calculated using the seeing of
the SINFONI observations, i.e 0.8 arcsec (0.8"$\times 10.4$~kpc$= 8320$ AU). Hence, as a function of the distance
$\dot{M}_{H_{2}}=3.8\times 10^{-9} (D/10.4$~kpc$)$M$_{\odot}$~yr$^{-1}$. The mass-loss rate estimated from H$_{2}$ emission lines would be only representative of
the total mass of the molecular component of the circumstellar gas swept up by a collimated high-velocity outflow. Low-luminosity and low-mass young stellar objects
are known to display similar features to those observed towards WF nebulae such as high-velocity, collimated outflows,
and shock-excited H$_{2}$ emission which leads to mass-loss rates estimates between $10^{-10}$ - $10^{-6}$ M$_{\odot}$~yr$^{-1}$ \citep[e.g.][]{Rebe13}.
The $\dot{M}_{H_{2}}$ value estimated for the WF IRAS 18043$-$2116 is within the H$_{2}$ mass-loss rate values that are usualy derived for these young stellar objects.
It could be that the energy input of the mechanism driving their high-velocity molecular outflows is similar to the case of Water Fountain nebulae. However,
observational evidence during the early post-AGB phase of (re-)accretion  of the material ejected during the AGB phase on the pre-White dwarf star
(or on a companion) is crucial to understand if accretion is related to the generation of the jet-driven winds observed towards WF nebulae.

If this is the case of an inhomogeneous wind, the mass-loss rate $\dot{M}_{ions}$ of ionized gas can be estimated assuming that the photoionization is maintained
with the kinetic energy of the wind, this using the ionizing photon rate calculated with Eq. \ref{eq:numberphot2}, and the Hydrogen ionization energy:
\begin{equation}
  \dot{N}_{i} h\nu = \frac{1}{2} \dot{M}_{\textrm{ions}} {\rm v}^{2}.
  \label{eq:kineticenergy}
\end{equation}
In the case of WF IRAS 18043$-$2116, we have several values related to the outflow velocity (v): The velocity of the most blue- and red-shifted spectral features, as well as
half the velocity range $\Delta$v at FWHM of the different H$_{2}$ transition. That is, 132~km~s$^{-1}$  and 264~km~s$^{-1}$ for the most blue- and red-shifted
maser spectral features, respectively, and 82~km~s$^{-1}$ as half the velocity range covered by the H$_{2}$ 1-0~S(1) emission at FWHM. Hence,
with v$= 132$~km~s$^{-1}$ and $h\nu=13.6$~eV, the mass-loss rate (as a function of the distance)
is $\dot{M}_{ions}= 3.3\times 10^{-5}~(D/10.4~\textrm{kpc})$~M$_{\odot}$yr$^{-1}$.
Using the LSR velocity of the molecular layers might lead to overstimate $\dot{M}_{ions}$ by not more than one order of magnitude, otherwise the total velocity
of the ionized outflow will be $>$1000~km~s$^{-1}$. In this case, $\dot{M}_{ions}$ should be representative of the ionized gas in the high-velocity outflow. In the case of
ionizing winds generating thermal radio continuum emission, \citet{Rey86} concluded that for a given observed radio flux, the mass-loss
rate inferred for jet-driven winds are considerable less than the values estimated if the ionizing winds are assumed to be spherical. According to that, if the radio
continuum emission detected is generated by a jet-driven wind, the mass-loss rate derived would be only a reliable upper limit of the ionized gas in the high-velocity outflow.

During the AGB phase, the mass-loss rates estimated from observations are typically in the range between $10^{-8}$~M$_{\odot}$yr$^{-1}$ - $10^{-4}$~M$_{\odot}$yr$^{-1}$, with
average expansion velocity of v$_{exp}\sim 15$~km~s$^{-1}$ \citep[e.g][]{Schoier,Sofia}. It is thought that such high mass-loss rates decrease to its minima due to the
exhaustion of the Hydrogen stellar atmosphere, once its mass is $\approx 10^{-2}$~M$_{\odot}$ \citep{AGBbook}. In the case of the WF IRAS 18043-2116, the velocity
field traced by the H$_{2}$ lines and H$_{2}$O maser emission is one order of magnitude larger than the average velocity measured towards the CSEs of AGB stars.
Therefore, although the mass-loss rate estimated for the high-velocity outflow and the values reported for the AGB wind are of the same order of magnitude,
is the velocity field traced by the spectral lines detected (H$_{2}$ and H$_{2}$O maser) towards the WF IRAS 18043-2116 what suggests that the mechanism driving
the high-velocity outflow is different from the mechanism driving the relic AGB wind.

Based on the spectral profile and on the full width at zero intensity (FWZI) of $^{12}$CO and $^{13}$CO lines detected towards a sample of evolved stars
(AGBs, post-AGBs and young Planetary nebulae), \citet{SanCon} estimated that the fraction of mass contained in the fastest outflows of their sample
(with $^{13}$CO FWZI$=$ 160~km~s$^{-1}$ and 120~km~s$^{-1}$, for IRAS 19374+2359 and IRAS 22036+5306 respectively) are 80\% and 55\% of the total CSE mass
associated to the detected CO emission. That means, bipolar structures with mass of 1.0~M$_{\odot}$ and 0.14~M$_{\odot}$ respectively. Although
the $\dot{M}_{ions}$ values estimated from the radio flux is an upper limit for IRAS 18043-2116 (the ionized outflow might be faster than the molecular layers),
fast outflows (v$_{lsr}>$100~km~s$^{-1}$) with mass-loss rates of the order of 10$^{-6}$-10$^{-4}$ M$_{\odot}$,
operating over timescales of the order of the post-AGB phase itself, can explain the formation of such high-mass bipolar outflows.

Therefore, if the WF IRAS 18043$-$2116 is a post-AGB star, and given the low mass hydrogen atmosphere of the pre-White Dwarf source, the
mass in the high-velocity outflows might be fed by a large scale structure different from the central post-AGB star. The accretion by the central star of
the material in a magnetized, rotating disk, is the most accepted scenario for the launching of collimated, high-velocity molecular outflows detected towards
young stellar objects in a wide mass range. In post-AGB stars, the formation of a rotating disk implies a perturbation on the spherically expanding AGB wind
scenario, mostly because the angular momentum of the material ejected from the AGB star is thought to be negligible. The formation of a rotating disk could
be well explained by the presence of a (sub-)stellar companion which provides angular momentum to the material ejected by the AGB star \citep{Zijlstra06}. Indeed, the
presence of disk structures in post-AGB stars has been inferred in most cases from Near- and/or Far-Infrared excess of emission, as well as from CO line emission, and
only a few of them have been spatially resolved with interferometric observations at submillimeter wavelegths \citep{Buja13,Buja16}.
Thus, we speculate that the presence of a magnetized, rotating disk could be the source of mass which is feeding the collimated, high-velocity outflow. High angular
resolution observations at submillimeter wavelengths would test our interpretation.

\section{Conclusions}

Although the source is not spatially resolved, the results from our multi-wavelength observations and the radiative transfer models indicate that:

\begin{itemize}
\item The Molecular emission detected, i.e the H$_{2}$O maser emission and the H$_{2}$ ro-vibrational lines
  emission, might be generated in the molecular layers around high-velocity outflows. The high velocity field traced by the molecular emission
  highlights the action of a mechanism that drives material with velocities at least one order of magnitude larger than the estimated average velocity field
  of the AGB wind.
\item Given the temporal gap of two years between the measurements of the flux density at 22~GHz taken with ATCA in 2013, and those at lower frequencies
  (1.5~GHz, 3~GHz, and 5~GHz) performed with JVLA in 2015, the best fits of the radiative transfer models suggest two possible scenarios:
  \begin{itemize}
  \item Thermal emission with a significant non-thermal (synchrotron) component: In this case, the flux density at 22~GHz is assumed to be constant between 2013 and 2015.
    The best fit suggests a non-thermal component with $\alpha=-0.7$, and flux density of 0.15~mJy at 22~GHz. Thus, a non-thermal component could be associated to
    the most energetic particles interacting with the ionizing shock front, whereas the thermal component could be associated to free particles with lower kinetic
    temperature at the post-shock region.
  \item Thermal emission: Our best fit indicates that, if the radio flux is dominated by thermal Bremsstrahlung emission, the flux density at 22~GHz should have
    increased about 20\% in the period between 2013 and 2015. From the density values retrieved from models that assume only thermal emission, the electron
    density should have increased 10\% in the two year period. Such growth of the density of thermal electrons could be associated to the propagation of a
    (partially) ionizing shock front throughout the slower AGB wind.
  \end{itemize}
\item The radio flux at the optically thin regime suggests mass-loss rates of the order of 10$^{-5}$~M$_{\odot}$yr$^{-1}$. Given the exhaustion of material on
  the stellar atmosphere of the mass-lossing star at the end of the AGB phase, we speculate that the outflowing material might be fed by another large scale
  structure, for instance, a circumstellar disk. High angular resolution observations are required in order to confirm both, the radio flux source, and whether the
  outflowing mass is channeled by (re-)accretion of material from a rotating circumtellar disk into the high-velocity outflow.
\end{itemize}

The results of our multi-wavelength observations support a scenario where the main energy input for the dynamics of large scale bipolar outflows
could be associated to both the launching mechanism, and to the propagation of a thermal radio jet throughout the CSE of the WF IRAS 18043$-$2116.
The detected ro-vibrational lines of molecular Hydrogen, high-velocity spectral features of H$_{2}$O maser emission detected at 22~GHz and 321~GHz, with
peak flux ratio $\approx 1$ \citep{Tafoya14}; radio flux at centimeter wavelegths, in addition to the previous reports on the detection OH maser lines, and in
particular of the OH 1720 transition \citep{Deacon04}, show the complexity of the processes ongoing in the CSE of this Water Fountain nebulae.

\begin{acknowledgements}
  Part of this work is based on observations collected at the European
  Organisation for Astronomical Research in the Southern Hemisphere
  under ESO programme 095.D-0574(A). The National Radio Astronomy Observatory is a
  facility of the National Science Foundation operated under cooperative agreement by Associated Universities, Inc.
  The Australia Telescope Compact Array is part of the
  Australia Telescope National Facility which is funded by the Australian Government for
  operation as a National Facility managed by CSIRO. AFPS aknowledges the Mexican Society
  of Physics and the National University of Mexico for postdoctoral
  fellowships during his residence in Mexico. RGL has received funding from the European
  Union’s Horizon 2020 research and innovation programme under the Marie Sklodowska-Curie grant
  agreement No 706320. WV and DT acknowledge support from ERC consolidator grant 614264.
\end{acknowledgements}

%
%

\bibliography{biblio}
\bibliographystyle{aa}
\end{document}